\documentclass[12pt,preprint,english]{aastex}
\usepackage[T1]{fontenc}
\usepackage[latin1]{inputenc}
\usepackage{color}
\usepackage{graphicx}
\usepackage{amssymb}
\usepackage{amsmath}
\usepackage{times}
\makeatletter

\providecommand{\tabularnewline}{\\}

\newcommand{\Ms}{M_{\star}}
\newcommand{\Rs}{R_{\star}}

\newcommand{\Mo}{M_{\odot}}

\newcommand{\Mbh}{M_{\bullet}}

\newcommand{\rMP}{r_{\mathrm{MP}}}

\begin{document}

%\centerline{\rule{\columnwidth}{1pt}}

\title{Massive perturbers in the galactic center }

\author{Hagai B. Perets \footnote{Faculty of Physics, Weizmann Institute of Science, POB 26, Rehovot 76100, Israel}, Clovis Hopman$^{1,}$\footnote{Leiden University, Leiden Observatory, P.O. box 9513, NL-2300 RA Leiden}  and Tal Alexander$^1$}

\begin{abstract}
We investigate the role of massive perturbers, such as stellar clusters or giant molecular clouds, in supplying low-angular momentum stars that pass very close to the central massive black hole (MBH) or fall into it. We show that massive perturbers can play an important role in supplying both binaries and single stars to the vicinity of the MBH. We discuss possible implications for the ejection of high velocity stars; for the capture of stars on tight orbits around the MBH; for the emission of gravitational waves from low-eccentricity inspiraling stars; and for the origin of the young main sequence B stars observed very near the Galactic MBH. Massive perturbers may also enhance the the growth rate of MBHs, and may accelerate the dynamical orbital decay of coalescing binary MBHs. 
\end{abstract}

\section{Introduction}
\label{s:intro}
There is compelling evidence that massive black holes (MBHs) lie in
the centers of all galaxies (e.g. \citep{Fer+00}), 
including the center of our Galaxy \citep{Eis+05,Ghe+05}. The MBH affects
the dynamics and evolution of the galaxy's center as a whole 
and it also strongly affects individual stars or
binaries that approach it. Such close encounters
 have been the focus of many studies and include a variety of processes 
such as 
destruction of stars by the MBH; capture and gradual inspiral
of stars into the MBH, accompanied by the emission of gravitational
waves (GWs); or dynamical
exchange interactions in which incoming stars or binaries energetically
eject a star tightly bound to the MBH and are captured in its place
very near the MBH \citep{Ale05}.  

The interest in such processes is driven by their possible implications
for the growth of MBHs, for the orbital decay of a MBH binary, for
the detection of MBHs, for GW astronomy, as well
as by observations of unusual stellar phenomena in our Galaxy, e.g.
the puzzling young population of B-star very near the Galactic MBH
, 
or the hyper-velocity B stars at the edge of the
Galaxy (e.g. \citep{Bro+05,Fue+06,Bro+06}), possibly ejected by 3-body interactions
of a binaries with the MBH.

Here we focus on close encounters with the MBH whose ultimate outcome
({}``event'') is the elimination of the incoming object from the
system, whether on the short infall (dynamical) time, if the event
is prompt (e.g. tidal disruption or 3-body exchange between a binary
and the MBH), or on the longer inspiral time, if the event progresses
via orbital decay (e.g. through GW emission).
Such processes are effective only when the incoming object follows
an almost zero angular momentum ({}``loss-cone'') orbit with periapse
closer to the MBH than some small distance $q$. To reach the MBH,
or to decay to a short period orbit, both the infall and inspiral
times must be much shorter than the system's relaxation time $t_{r}$. 
The fraction of stars initially on loss-cone orbits
is very small and they are rapidly eliminated. Subsequently, the close
encounter event rate is set by the dynamical processes that refill
the loss-cone.

The loss-cone formalism used for estimating the event rate (e.g.\citep{Coh+78})
usually assumes that the system is isolated and that the refilling
process is 2-body relaxation. This typically leads to a low event
rate, set by the long 2-body relaxation time.

Two-body relaxation, which is inherent to stellar systems, ensures
a minimal loss-cone refilling rate. Other, more efficient but less
general refilling mechanisms were also studied with the aim of explaining
various open questions,
or in the hope that they may lead to significantly higher event rates
for close encounter processes \citep{Mer06}. 
However, most of
these mechanisms require special circumstances to work, or are short-lived. 

Here we explore another possibility, which is more likely to apply
generally: accelerated relaxation and enhanced rates of close encounters
driven by massive perturbers (MPs). Efficient relaxation by MPs were
first suggested in this context by Zhao, Haehnelt \& Rees \citep{Zha+02} 
as a mechanism
for establishing the $M_{\bullet}/\sigma$ relation \citep{Fer+00}
by fast accretion of stars and dark matter. Zhao et al. also noted
the possibility of increased tidal disruption flares and accelerated
MBH binary coalescence due to MPs. In this study we investigate in
detail the dynamical implications of relaxation by MPs. We evaluate
its effects on the different modes of close interactions with the
MBH, in particular 3-body exchanges, which were not considered by
Zhao et al. and apply our results to the Galactic Center (GC),
where observations indicate that dynamical relaxation is very likely
dominated by MPs.

\section{Loss-cone refilling }
\label{s:outline}
In addition to stars, galaxies contain dense massive structures
 such as molecular clouds, open clusters and globular clusters with
masses up to $10^{4}$--$10^{7}\,\Mo$. Such structures can perturb
stellar orbits around the MBH much faster than 2-body stellar relaxation
(hereafter {}``stellar relaxation''), provided they are numerous
enough. 
The minimal impact parameter
still consistent with a small angle deflection in the MP-star
scattering is $b_{\min}\!=\! GM_{p}/v^{2}$
(the capture radius), where $v$ is of the order of the local velocity
dispersion $\sigma$. 
%Defining $B\!\equiv\! b/b_{\min}\!\ge\!1$,
%the encounter rate is then \begin{align}
%\left(\frac{\mathrm{d}^{2}\Gamma}{\mathrm{d}M_{p}db}\right)\mathrm{d}M_{p}\mathrm{d}b\, & \sim\left(\frac{\mathrm{d}N_{p}}{\mathrm{d}M_{p}}\right)\mathrm{d}M_{p}vb_{\min}^{2}2\pi B\mathrm{d}B =\frac{G^{2}}{v^{3}}\left[\left(\frac{\mathrm{d}N_{p}}{\mathrm{d}M_{p}}\right)M_{p}^{2}\right]\mathrm{d}M_{p}2\pi B\mathrm{d}B\,.\end{align}
 The total rate of scattering stars into the loss cone, $\Gamma$,  is obtained 
by integrating $d\Gamma/dM_pdb$ over all MP masses 
and
over all impact parameters between $b_{\min}$ and $b_{\max}\approx r$. 
The relaxation rate due to all MPs at all impact parameters is then
\begin{align}
t_{r}^{-1} & =\int_{b_{\min}}^{b_{\max}}\mathrm{d}b\mathrm{\int}\mathrm{d}M_{p}\left(\frac{\mathrm{d}^{2}\Gamma}{\mathrm{d}M_{p}db}\right) \sim\log\Lambda\frac{G^{2}}{v^{3}}\int\mathrm{d}M_{p}\left(\frac{\mathrm{d}N_{p}}{\mathrm{d}M_{p}}\right)M_{p}^{2}\,,\end{align}
where  $\log\Lambda\!=\!\log(b_{\max}/b_{\min}$) is the Coulomb logarithm
(here the dependence of $\log\Lambda$ and $v$ on $M_{p}$ is assumed
to be negligible). 
This formulation of the relaxation time is equivalent to its conventional
definition  \citep{Bin+87} as the time for a change of order unity
in $v^{2}$ by diffusion in phase space due to scattering, $t_{r}\!\sim\! v^{2}/D(v^{2})$,
where $D(v^{2})$ is the diffusion coefficient. 
If the stars and MPs have distinct mass scales with typical number
densities \textcolor{black}{$N_{\star}$ and $N_{p}$ and rms masses}
$\left\langle M_{\star}^{2}\right\rangle ^{1/2}$ and $\left\langle M_{p}^{2}\right\rangle ^{1/2}$
(\textcolor{black}{$\left\langle M^{2}\right\rangle \!\equiv\!\int M^{2}(\mathrm{d}N/\mathrm{d}M)\mathrm{d}M/N$})\textcolor{black}{,
then MPs dominate if} the ratio of the 2nd moments of the MP and star
mass distributions, $\mu_{2}\!\equiv\!\left.N_{p}\left\langle M_{p}^{2}\right\rangle \right/N_{\star}\left\langle M_{\star}^{2}\right\rangle $,
satisfies $\mu_{2}\!\gg\!1$.

\begin{figure}[h]
\includegraphics[width=14pc]{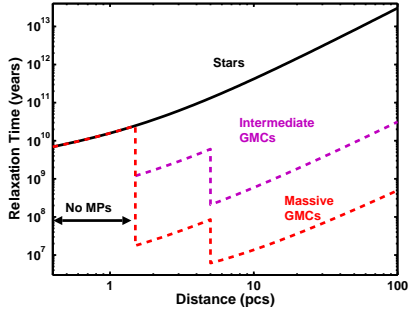}
\hspace{2pc}
\begin{minipage}[b]{19pc}
\caption{\label{f:tr} {\footnotesize Relaxation time as function of distance from the MBH,
for stars (solid line) and for each of the 2 MP models separately,
as listed in table \ref{t:models}, massive GMCs (dashed-dotted line)
and intermediate GMCs (dashed line). 
The discontinuities are artifacts of the assumed
sharp spatial cutoffs on the MP distributions. 2-body stellar processes
dominate close to the MBH, where no MPs are observed to exist. However,
at larger distances massive clumps (at $1.5<r<5$ pcs) and GMCs (at
$5<r<100$) are much more important.} }
\end{minipage}
\end{figure}

The central $\sim\!100\,\mathrm{pc}$
of the GC contain $10^{8}-10^{9}$ solar masses
in stars, and about $10^{6}-10^{8}$ solar masses in MPs such as open
clusters and GMCs of masses $10^{3}-10^{7}\,\Mo$ \citep{Oka+01,Fig+02,Vol+03,Gus+04,Bor+05}.
An order of magnitude estimate indicates that MPs in the GC can reduce
the relaxation time by several orders of magnitude, \begin{align}
\frac{t_{r,\star}}{t_{r,\mathrm{MP}}} & =\mu_{2}\sim\frac{(N_{p}M_{p})M_{p}}{(N_{\star}M_{\star})M_{\star}} =10^{3}\left[\frac{(N_{\star}\Ms/N_{p}M_{p})}{10^{2}}\right]^{-1}\left[\frac{(M_{p}/\Ms)}{10^{5}}\right]\,.\end{align}
 This estimate is borne by more detailed calculations (Fig. \ref{f:tr}
and table \ref{t:models}), using the formal definition $t_{r}\!=\! v^{2}/D(v_{||}^{2})$
with $M_{\star}\rho_{\star}\rightarrow\int(\mathrm{d}N_{p}/\mathrm{d}M_{p})M_{p}^{2}\mathrm{d}M_{p}$
\citep{Bin+87}. 
In our calculations we follow the Fokker-Planck approach to the loss-cone problem \citep{Coh+78}, where we recalculate the diffusion coefficients obtained
from stellar two body relaxation \citep{Bin+87} 
by taking into account the contribution from 
the much more massive MPs (for details see \citep{Per+06}). 

Although some of the assumptions concerning the loss cone formalism are not necessarily valid in the case of MPs, 
the loss-cone formalism can be generalized to deal with MPs in an
approximate manner with only few modifications \citep{Per+06}.
Thus to a good approximation our calculation follows the usual loss
cone treatment (e.g. \citep{Mag+99}) where the 
mass of the stars
is replaced by that of the MPs, and the integration over the energies
of deflected stars (mapped to their distance from the MBH) is done
only for regions where MPs exist.
\section{Massive perturbers in the Galactic Center}
\label{s:GC_MPs}
MPs can dominate relaxation only when they are massive enough to compensate
for their small space densities. Here we consider only MPs with masses
$M_{p}\!\ge\!10^{3}M_{\odot}$. Such MPs could be open or globular
stellar clusters or molecular clouds of different masses, in particular
giant molecular clouds (GMCs). Observations 
of the Galaxy reveal enough MPs to dominate  relaxation
in the central 100 pc. We adopt here a conservative approach, and
include in our modeling only those MPs that are directly observed
in the Galaxy. Based on the observations
of such MPs we devise several possible models for the MPs in the GC, 
detailed in table \ref{t:models}.
As these observations \citep{Oka+01,Fig+02,Vol+03,Gus+04,Bor+05} show 
that the MPs population is dominated by the GMCs, we consider only GMCs and gaseous clumps in our models.
The observed MP species vary in their spatial distributions and mass
functions, which are not smooth or regular. For our numeric calculations,
we construct several simplified MP models (table \ref{t:models})
that are broadly based on the observed properties of the MPs, and assume that the spatial distribution of MPs follows that of the stars (apart of the inner radius cutoff where no MPs are observed).

The three MP models in table \ref{t:models}: Stars, GMC1 and GMC2, represent respectively
the case of relaxation by stars only,  by heavy
GMCs and light GMCs.
\begin{table}
\begin{center}
\caption{\label{t:models}Massive perturber models }
\begin{tabular}{ccccrcc}
\hline 
\multicolumn{1}{c}{Model}&
$r$ (pc)$\,{}^{a}$&
$N_{p}$&
 $M_{p}$ ($M_{\odot}$)&
$\beta$&
$R_{p}$ (pc)&
$\mu_{2}\,^{b}$\tabularnewline
\hline
GMC1&
5--100 &
$100$&
$10^{4}\!-\!10^{7}$&
$1.6$&
5&
$3\!\times\!10^{5}$\tabularnewline
&
1.5--5&
$30$&
$10^{3}\!-\!10^{5}$&
$0.9$&
&
$6000$\tabularnewline
GMC2&
5--100 &
$100$&
$10^{3}\!-\!10^{6}$&
$1.6$&
5&
$3\!\times\!10^{3}$\tabularnewline
&
1.5--5&
$30$&
$10^{2}\!-\!10^{4}$&
$0.9$&
&
$60$
\tabularnewline
Stars&
5--100 &
$2\!\times\!10^{8}$&
$1$&
---&
$\sim\!0$&
1\tabularnewline
Stars&
1.5--5 &
$6\!\times\!10^{6}$&
$1$&
---&
$\sim\!0$&
1\tabularnewline
\hline
\multicolumn{7}{l}{{\footnotesize $^{a}$ $N_{p}(r)\!\propto\! r^{-2}$
assumed.}}\tabularnewline
\multicolumn{7}{l}{{\footnotesize $^{b}\,$ $^{b}$} $\mu_{2}\!\equiv\! N_{p}\left\langle M_{p}^{2}\right\rangle \left/N_{\star}\left\langle \Ms^{2}\right\rangle \right.$,
where \textcolor{black}{\footnotesize $\left\langle M^{2}\right\rangle \!=\!\int M^{2}(\mathrm{d}N/\mathrm{d}M)\mathrm{d}M/N$.}}\tabularnewline
\hline
\end{tabular}\par
\end{center}
\end{table}

\section{Massive perturber-driven interactions with a MBH}
\label{s:interact}
The maximal differential loss-cone refilling rate, which is also the
close encounters event rate, $\mathrm{d}\Gamma/\mathrm{d}E$, is reached
when relaxation is efficient enough to completely refill the loss
cone during one orbit. 
Further decrease in the
relaxation time does not affect the event rate at that energy. MPs
can therefore increase the differential event rate over that predicted
by stellar relaxation, only at high enough energies, $E\!>\! E_{c}$
(equivalently, small enough typical radii, $r\!<\! r_{c}$), the critical 
energy (radius), separating the full and empty loss cone regimes, where
slow stellar relaxation fails to refill the empty loss-cone. The extent
of the empty loss-cone region increases with the maximal periapse
$q$, which in turn depends on the close encounter process of interest.
For example, the tidal disruption of an object of mass $M$ and size
$R$ occurs when $q\!<\! r_{t}$,  the tidal disruption radius, $ r_{t}\simeq R\left(\Mbh/M\right)^{1/3}\,.$
This approximate disruption criterion applies both for single stars
($M\!=\!\Ms$, $R\!=\!\Rs$) and for binaries,  where  $M$ is the
combined mass of the binary components and $R$ is the binary's semi-major
axis, $a$. Stellar radii are usually much smaller than typical binary
separations, but stellar masses are only $\sim\!2$ times smaller
than binary masses. Binaries are therefore disrupted on larger scales
than single stars. In the GC this translates to an empty (stellar
relaxation) loss-cone region extending out to $r_{c}^{s}\!\sim\!3$
pc for single stars and out to $r_{c}^{b}\!>\!100$ pc for binaries.
In the GC $\rMP\!\lesssim\! r_{c}^{s}\!\ll\! r_{c}^{b}$ (where $\rMP$
 is the smallest distance from the MBH where MPs are observed) , and so
MPs are expected to increase the binary disruption rate by orders
of magnitude, but increase the single star disruption rate only by
a small factor. 
Since for most Galactic MP types $\rMP\!>\! r_{h}$ (the radius of influence of the MBH), the disruption
rate is dominated by stars near $r_{\mathrm{MP}}$. For example, when
the loss-cone is empty, $\sim\!50\%$ of the total rate is due to
MPs at $r\!<\!2\rMP$; when the loss-cone is full, $\sim\!75\%$ of
the total rate is due to MPs at $r\!<\!2\rMP$ (see details in \citep{Per+06}). 

\subsection{Interactions with single stars}
\label{ss:single}
Clusters, GMCs and gas clumps in the GC are abundant only beyond the
central $r_{\mathrm{MP}}\!\sim\!1.5$ pc, whereas the empty loss-cone
regime for tidal disruption of single stars extends only out to $r_{c}^{s}\!\sim\!3$
pc. For inspiral processes such as GW emission, $r_{c}$ is $\sim\!100$
times smaller still \citep{Hop+05}. The effect of such MPs
on close encounter events involving single stars is thus suppressed
(weaker tidal effects by MPs at $r\!>\! r_{c}^{s}$ are not considered
here). This is contrary to the suggestion of Zhao et al. \citep{Zha+02}, who
assumed that the effect of MPs fully extends to the empty loss-cone
regime. We find that the enhancement of MPs over stellar relaxation
to the single stars disruption rate  is small, less than a factor
of $3$, and is due to stars scattered by gas clumps in the small
empty-loss cone region between $\rMP\!\sim\!1.5\,\mathrm{pc}$ and
$r_{c}^{s}\!\sim\!3\,\mathrm{pc}$. A possible exception to this conclusion
is the possible existence of IMBHs population,  not modeled here.
\subsection{Interactions with stellar binaries}
\label{ss:binary}
The empty loss cone region for binary-MBH interactions extends out
to $>\!100$ pc because of their large tidal radius. On these large
scales MPs are abundant enough to dominate the relaxation processes.
Here we focus on 3-body exchange interactions \citep{Hil88,YuQ+03},
which lead to the disruption of the binary, the energetic ejection
of one star, and the capture of the other star on a close orbit around
the MBH. The event rate is highly dependent on the unknown binary fraction 
in these regions.

The binary fraction and typical binary semi-major axis depend on the
binary mass, and on the rate at which binaries evaporate by encounters
with other stars. This depends in turn on the stellar densities and
velocities, and therefore on the distance from the MBH. We take
these factors into account and estimate in detail the 3-body exchange
rate for MP-driven relaxation. The rate is proportional to the binary
fraction in the population, which is the product of the poorly-known
binary IMF in the GC and the survival probability against binary evaporation. 
 
The capture probability and the semi-major axis distribution of captured
stars were estimated by simulations \citep{Hil91,YuQ+03}. Numeric
experiments indicate that between $0.5$--$1.0$ of the binaries that
approach the MBH within the tidal radius $r_{t}(a)$ 
are disrupted. Here we adopt a disruption efficiency of $0.75$. The
harmonic mean semi-major axis for 3-body exchanges with equal mass
binaries was found to be \citep{Hil91} \begin{equation}
\left\langle a_{1}\right\rangle \simeq0.56\left(\frac{M_{\bullet}}{M_{\mathrm{bin}}}\right)^{2/3}a\simeq\,0.56\left(\frac{M_{\bullet}}{M_{\mathrm{bin}}}\right)^{1/3}r_{t},\label{e:afinal}\end{equation}
 where $a$ is the  semi-major axis of the infalling binary and $a_{1}$
that of the captured star (the MBH-star {}``binary''). Most values
of $a_{1}$ fall within a factor $2$ of the mean. This relation maps
the semi-major axis distribution of the infalling binaries to that
of the captured stars: the harder the binaries, the more tightly bound
the captured stars. 
The periapse of the captured star is at $r_{t}$,
and therefore its eccentricity is very high \citep{Hil91,Mil+05},
$e\!=\!1-r_{t}/a_{1}\!\simeq1-1.8(M_{\mathrm{bin}}/M_{\bullet})^{1/3}\!\gtrsim\!0.98$
for values typical of the GC. 

We now consider the implications of 3-body exchange interactions
of the MBH with old ($t_{\star}\!\gtrsim\! t_{H}$) binaries and massive
young ($t_{\star}\!<\!5\!\times10^{7}$ yr) binaries.

The properties of binaries in the inner GC are at present poorly determined.
We use the period distribution of Solar neighborhood binaries for old low mass binaries \citep{Duq+91} and young massive binaries \citep{Kob+06}. 
The total binary fraction of these binaries is
estimated at $f_{\mathrm{bin}}\sim\!0.3$ for low mass \citep{Lad06} 
binaries, and $f_{\mathrm{bin}}\sim\!0.75$ for  massive binaries \citep{Kob+06}. Adopting
these values for the GC, the total binary disruption rate by the MBH
can  then be calculated by integrating
$\mathrm{d}N_{\mathrm{bin}}/\mathrm{d}a$ %(Eqs. \ref{e:fhard}) 
over the binary $a$ distribution  and over  the power-law stellar density
distribution of the GC from the minimal radius where such binaries exist 
(for young binaries 
we assume an inner cutoff at 1.5 pcs, where such young stars are not observed
\citep{Pau+06}), up to 100 pc \citep{Gen+03}. Table (\ref{t:bin})
lists the capture rates for the different perturber models, assuming
a typical old equal-mass binary of $M_{\mathrm{bin}}\!=\!2\,\Mo$, or young
equal-mass binary of $M_{\mathrm{bin}}\!=\!15\,\Mo$.

The old, low-mass binary disruption rate we derive for stellar relaxation
alone is $\sim5\times10^{-7}\,\mathrm{yr^{-1}}$, $\sim\!5$ times
lower, but still in broad agreement with the result of Yu \& Tremaine \citep{YuQ+03}.
Their rate is somewhat higher because they assumed a constant binary
fraction and a constant semi-major axis for all binaries, even  close
to the MBH, where these assumptions  no longer hold. 

MPs increase the binary disruption and high-velocity star ejection
rates by factors of $\sim\!10^{1-3}$ and effectively accelerate stellar
migration to the center. 
Thus we expect an increase in the number of stars captured close to the MBH 
and consequently a higher event rate of single
star processes such as tidal disruption, tidal heating and GW emission
from compact objects, in particular from compact objects on zero-eccentricity
orbits \citep{Mil+05}.
\begin{table}
\begin{center}
\caption{\label{t:bin}Total binary disruption rate and number of captured
young stars }
\begin{tabular}{ccccc}
\hline 
Model&
\multicolumn{2}{c}{Disruption rate ($\mathrm{yr^{-1}}$) }&
Young Stars$^{a}$&
Young Stars$^{a}$ \tabularnewline
&
$r\!<\!0.04\,\mathrm{pc}$&
$r\!<\!0.4\,\mathrm{pc}$&
$r\!<\!0.04\,\mathrm{pc}$&
$0.04\,<\, r\!<\!0.4\,\mathrm{pc}$\tabularnewline
\hline
GMC1&
$1\times10^{-4}$&
$2.8\times10^{-4}$&
$33.5$&
$3.8$\tabularnewline
GMC2&
$2.1\times10^{-5}$&
$6.4\times10^{-5}$&
$5$&
$0.24$
\tabularnewline
Stars&
$3.4\times10^{-7}$&
$5.3\times10^{-7}$&
$0.15$&
$0.003$\tabularnewline
Observed&
?&
?&
$10-35^{b}$&
$?$\tabularnewline
\hline
\multicolumn{5}{l}{{\footnotesize $^{a}$Main sequence B stars with lifespan $t\!<\!5\!\times\!10^{7}\,\mathrm{yr}$
.}}\tabularnewline
\multicolumn{5}{l}{{\footnotesize $^{b}\,$ $\sim\!10$ stars with derived $a\!\lesssim0.04\,\mathrm{pc}$.
$\gtrsim\!30$ stars are observed in the area.}}\tabularnewline
\hline
\end{tabular}
\end{center}
\end{table}
MPs may be implicated in the puzzling presence of a cluster of main
sequence B-stars ($4\!\lesssim\!\Ms\!\lesssim\!15\,\Mo$) in the inner
$\sim\!1^{''}$ ($\sim\!0.04$ pc) of the GC. This so-called {}``S-cluster''
is spatially, kinematically and spectroscopically distinct from the
young, more massive stars observed farther out, on the $\sim\!0.05$--$0.5$
pc scale, which are thought to have formed from the gravitational
fragmentation of one or two gas disks \citep{Pau+06}.
There is however still no satisfactory explanation for the existence
of the seemingly normal, young massive main sequence stars of the
S-cluster, so close to a MBH (see review of proposed models by Alexander \citep{Ale05};
also a recent model by Levin \citep{Lev06}, and in these proceedings). 

Here we revisit an  idea proposed by Gould \& Quillen \citep{Gou+03},
 that the S-stars
were captured near the MBH by 3-body exchange interactions with infalling
massive binaries. Originally, this exchange scenario lacked a plausible
source for the massive binaries. 
We suggest that MP-driven 3-body exchanges can serve as a source
for binaries, as they
increase the rate of infall of young field binaries to the MBH. 
Such young field binaries
should exist in the GC, taking into account
the ongoing star formation in the central
$\sim\!100$ pc, which implies the presence of a large reservoir of massive
stars there. Such stars are indeed  observed in the central $\mathrm{few}\times10$
pc both in dense clusters and in the field \citep{Fig+02,Mun+06}.
It is plausible that a high fraction of them are in binaries.

We assume star formation at a constant rate for 10 Gyr with a Miller-Scalo
IMF \citep{Mil+79}, and use a stellar population synthesis code \citep{Ste+03}
with the Geneva stellar evolution tracks \citep{Sch+92a} to estimate
that the present day number fraction of stars in the S-star mass range
is  $3.5\!\times\!10^{-4}$ (and less than 0.01 of that for $\Ms\!>\!15\,\Mo$
stars). Note that if star formation in the GC is biased toward massive
stars \citep{St0+05}, this estimate should be revised upward.
We adopt the observed Solar neighborhood distribution of the semi-major
axis of massive binaries, which shows that 
massive binaries are thus typically harder than low-mass binaries \citep{Kob+06},
and will be tidally disrupted 
closer to the MBH
and leave a more tightly bound captured star.

We represent the massive binaries by one with equal mass stars in
the mid-range of the S-stars masses,  with $M_{\mathrm{bin}}\!=\!15\,\Mo$
and $t_{\star}(7.5\,\Mo)\simeq5\!\times\!10^{7}\,\mathrm{yr}$, and
integrate over the stellar distribution and the binary $a$ distribution
as before, to obtain the rate of binary disruptions, $\Gamma$, the
mean number of captured massive stars in steady state, $N_{\star}\!=\!\Gamma t_{\star}$,
and their semi-major axis distribution (Eq. \ref{e:afinal}). Table
(\ref{t:bin}) compares the number of captured young stars in steady
state, for the different MP models, on the $r\!<\!0.04\,\mathrm{pc}$
scale (the S-cluster) and $0.04\!<\! r\!<\!0.4$ pc scale (the stellar
rings) with current observations \citep{Eis+05,Pau+06}. 

\begin{figure}[h]
\includegraphics[width=15pc]{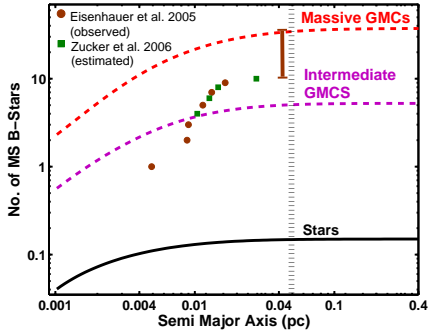}
\hspace{2pc}
\begin{minipage}[b]{19pc}
\caption{\label{f:Scluster} Cumulative number of young B-stars in the GC as
predicted by the MP models and by stellar two-body relaxation (listed
in table 2).The circles, squares and vertical bar represents the cumulative
number 
of observed
young stars inside 0.04 pc \citep{Eis+05,Zuc+06} . The dotted vertical line marks the approximate maximal distance
in which captured B-stars are expected to be observed. }
\end{minipage}
\end{figure}

The number of captured massive stars falls rapidly beyond $0.04$ pc 
(table \ref{t:bin}) where the S-cluster is observed
because wide massive binaries are rare. This capture model thus provides
a natural explanation for the central concentration of the S-cluster
(Fig \ref{f:Scluster}). The absence of more massive stars in the
S-cluster ($\Ms\!>\!15\,\Mo$, spectral type O\,V) is a statistical
reflection of their much smaller fraction in the binary population.
Figure (\ref{f:Scluster}) and table (\ref{t:bin}) compare the cumulative
semi-major axis distribution of captured B-stars, as predicted by
the different MP models, with the total number of young stars observed
in the inner 0.04 pc ($\sim\!35$ stars \citep{Eis+05,Ghe+05,Pau+06}.
Of these, only $\sim\!10$ have full orbital solutions (in particular
$a$ and $e$) at present. 
The numbers
predicted by the MP models are consistent with the observations, unlike
the stellar relaxation model that falls short by two orders of magnitude.

The binary capture model predicts that captured stars have very high
initial eccentricities. Most of the solved S-star orbits do have $e\!>\!0.9$,
but a couple have $e\!\sim\!0.3$--$0.4$  \citep{Eis+05}. Normal stellar relaxation is too slow to explain the decrease in the
eccentricity of these stars over their relatively short lifetimes.
However, the much faster process of resonant relaxation \citep{Rau+96}
may be efficient enough to randomize the eccentricity of a fraction
of the stars, and could thus possibly explain the much larger observed
spread in eccentricities \citep{Hop+06}. 

The companions of these captured stars are ejected from the GC
 at high velocities \citep{Hil88}. Consequently, our model predicts the number of young 
hyper velocity stars ejected from the GC (such as observed \citep{Bro+05,Fue+06,Bro+06}) to be comparable to that of observed B-stars in the 
S-cluster, in agreement with current observation based estimations \citep{Bro+05,Fue+06,Bro+06}.

\section{Summary}
\label{s:summary}
We presented here the results of a study of the effect of massive 
perturbers near
the MBH in the GC. We have shown that current observations of MPs
such as GMCs and clusters indicate that they dominate relaxation process
in the GC, where they exist, and they are much more important than stellar
two-body relaxation processes by stars which are usually considered.
We have used the loss cone formalism in order to analyze the importance
of this mechanism in generating low angular momentum stars and binaries
to interact with the MBH. We have have computed the rates of these
interactions, and showed that they can be highly important for the 
process of binaries
disruption by the MBH and its consequent outcomes. The origin of the
young massive B-stars at the central arcsecond of the GC can be explained
by the capture of stars from binaries disrupted by the MBH after being
scattered by MPs. Our calculations also show that some of the companions
of these captured stars could be ejected at high velocities, thus
explaining the observations of young massive high velocity stars observed
recently. In addition MPs increase the number of compact stars captured
close to the MBH, and thus increase the emission of zero-eccentricity
GWs. We also suggest that such MPs may also help solve the
last parsec problem of coalescing BMBHs. Although we focused
on the GC of the milky-way, our results can be easily extended to
MPs near other MBHs.

% Add if there is enough space
%\ack We thank the organizers of the GC2006 meeting for a very
%enjoyable and stimulating conference.

\end{document}